\documentclass[12pt,journal]{IEEEtran}
\usepackage{array}
\usepackage{supertabular}
\usepackage{graphicx}
\usepackage{caption}
\usepackage{amsmath}
\usepackage{bm}
\usepackage{amsfonts}
\usepackage{xspace}
\usepackage{floatrow,subfig}
\usepackage{url}
\usepackage{balance}

\begin{document}
\newcommand{\Ly}{Lyapunov\xspace}
%
\title{Energy Trading between Microgrids  Individual Cost Minimization and Social Welfare Maximization}
%
%
%
%

\author{{~Zhenyu~Qiao,~Bo~Yang,~Qimin~Xu,~Fei~Xiong,~Cailian~Chen,~Xinping~Guan,~Bei~Chen}
\IEEEcompsocitemizethanks{\IEEEcompsocthanksitem Z. Qiao, B. Yang, C. Chen, Q. Xu, F. Xiong, C. Chen and X. Guan are with the Department of Automation, Shanghai Jiao Tong University, Shanghai, 200240 P. R.
\IEEEcompsocthanksitem Z. Qiao, B. Yang, C. Chen, Q. Xu, F. Xiong, C. Chen and X. Guan are also with Collaborative Innovation Center for Advanced Ship and Deep-Sea Exploration, Shanghai, China and Key Laboratory of System Control and Information Processing, Ministry of Education of China, Shanghai, China
\IEEEcompsocthanksitem B. Chen is with Shanghai Electric Group Co., Ltd Central Academe, Shanghai, China.}}
\IEEEcompsoctitleabstractindextext{%
\begin{abstract}
High penetration of renewable energy source makes microgrid (MGs) be environment friendly. However, the stochastic input from renewable energy resource brings difficulty in balancing the energy supply and demand. Purchasing extra energy from macrogrid to deal with energy shortage will increase MG energy cost. To mitigate intermittent nature of renewable energy, energy trading and energy storage which can exploit diversity of renewable energy generation across space and time are efficient and cost-effective methods. But current energy storage control action will impact the future control action which brings challenge to energy management. In addition, due to MG participating energy trading as prosumer, it calls for an efficient trading mechanism. Therefore, this paper focuses on the problem of MG energy management and trading. Energy trading problem is formulated as a stochastic optimization one with both individual profit and social welfare maximization. Firstly a \Ly optimization based algorithm is developed to solve the stochastic problem. Secondly the double-auction based mechanism is provided to attract MGs' truthful bidding for buying and selling energy. Through theoretical analysis, we demonstrate that individual MG can achieve a time average energy cost close to offline optimum with tradeoff between storage capacity and energy trading cost. Meanwhile the social welfare is also asymptotically maximized under double auction. Simulation results based on real world data show the effectiveness of our algorithm.
\end{abstract}

}

\maketitle
\IEEEdisplaynotcompsoctitleabstractindextext
\IEEEpeerreviewmaketitle

\section{Introduction}
\IEEEPARstart{G}{rowing} attentions to the environment and demands of electricity promote the power grid modernizing. A microgrid (MG) is a distributed system equipped with renewable energy sources which can reduce greenhouse gas emission. However stochastic variations of energy produced by renewable energy sources make MG be hard to achieve balance between energy supply and demand. The risk of single MG to meet certain operational objective (e.g. minimizing energy costs) is very high.  A popular technique to compensate for stochastic renewable energy generation is energy storage which can buff energy and copy with future energy shortage. However, with deeper integration of renewable energy resources, increasing storage capacity is no longer a cost-effective method. \par
As a result, the need for more cost-effective solutions has motivated studies of energy trading between interconnected MGs \cite{and}. Geographical distributed MGs have chance to increase overall profits by trading due to diversity of energy generation. Nevertheless MGs are energy providers and consumers. Even more MG needs to make some important decisions: whether use more energy to serve their own loads, energy charging or trading? How much money should be charged for per unit energy? These decisions and question are critical to the economics of MG. These decisions should be efficiently and optimally made in an online fashion, which can help MG to get profit from trading without decreasing user experience and guarantee long-term optimality of individual MG profit, as well as the social welfare.\par
In this paper, we deal with MG energy management and trading problem considering both stochastic renewable energy and random power demands. First, we present a Lyapunov-optimization based algorithm to help MG to calculate prices and amounts of energy for trading. \Ly-optimization can achieve a result which is close to offline optimum without specific probability distribution of renewable energy generation and customer demands. Then a double-auction based trading mechanism is proposed for interconnected MGs energy trading in which no MG can be benefit from untruthful bidding. In particular, our main contributions of this paper are as follows:
\begin{itemize}
\item From the perspective of MGs, we model a comprehensive MG mainly including renewable energy and energy storage in a energy trading market. Then we propose a two-stage energy trading mechanism which is economic efficient, individual rational, balanced budge and truthfulness to minimize the individual cost of MG and maximize social welfare by energy trading.
\item We combine a double auction scheme with stochastic \Ly optimization based algorithm for interconnected MGs trading. Each MG can price energy based on the current status of customer demands and energy storage without future renewable energy generation information. Thus double auction mechanism can guarantee more profits than that when MG operates alone.
\item Through theoretical analysis, our algorithm can achieve better trade-off among energy trading, energy storage capacity and customers dissatisfaction index. Additionally, energy trading will not cause profit  loss of MG. Moreover, by using practical data sets, we prove the effectiveness of our proposed algorithm.
\end{itemize}

In the remainder of the paper, we discuss the system model and double auction framework in Section III. Then, in Section IV, we propose a solution based on \Ly optimization for the former problem,and prove the theoretical performance of this method. Numerical results are in Section V and conclusion is in the Section VI.

\section{Related Works}
There has been some recent researches in energy trading between MGs. References \cite{economic} and \cite{power} consider an individual MG operation with energy storage and renewable energy source. However, MG operated without cooperation will waste some of renewable energy due to limit energy storage capacity. In \cite{and}, cooperation allows MG to borrow energy from other MGs which have extra energy. Iterative double auction is introduced into trading in \cite{iterative} and \cite{game-theoretic}. Auction mechanism ensures the participants benefit from cooperation and trading when auctioneer get complete information of MG current status. Reference \cite{virtual} adjusts double auction mechanism for another scenario called inter-cloud trading. The algorithm can reach asymptotical social welfare maximization with limited information which can protect individual cloud privacy. Nevertheless, a energy trading framework applied within MGs should be developed. MG will attend the trading with considering renewable energy, energy storage and customer demands. \par

\section{System Model and Double Auction Framework}
\setcounter{figure}{0}
\subsection{System Model}
\begin{figure}
\includegraphics[totalheight=60mm,width=80mm]{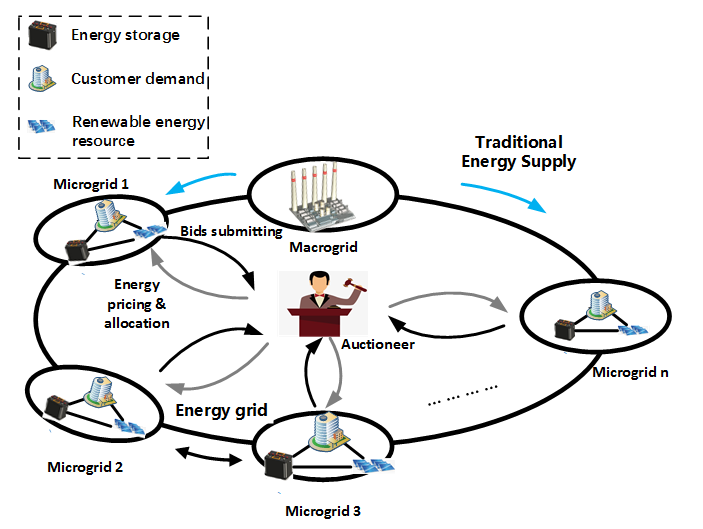}
\caption{Energy trading within interconnected microgrids through an auction process}
\end{figure}
We consider an inter-connected power grid consisting of $n$ microgrids (MGs) and a macro-grid as shown in Fig. 1. The MGs are capable of harvesting renewable energy (e.g. wind, solar energy etc.). In addition, the MGs are equipped with energy storage in which they can store the harvested energy for future use.
\subsubsection{Energy Generation and Purchase}
$\mbox{MG}_{i}$ harvests $R_{i}(t)$ units of energy during one time slot $t$. One time slot typically is half or one hour, we assume one time slot as one hour in order to coordinate with simulation. Renewable energy is first stored in the energy storage before it can be used in next time slot. The macrogrid generates the energy in the traditional way. We assume that the macrogrid has a very large energy generation in one time slot (which means we do not need to consider any macrogrid energy generation constraints). $\mbox{MG}_{i}$ purchases $G_{i}(t)$ units of energy from macrogrid with price $P(t)$.
\subsubsection{Energy Storage}
\setcounter{equation}{0}
Each MG has storage which can store extra energy. The energy in the storage can be considered as a queue $B_{i}(t)$. Renewable energy used for charging denote as $C_{i}(t)$ which means energy enter the storage queue, meanwhile discharging $D_{i}(t)$ means the energy leaving the storage queue. Then we can have following equation:
\begin{equation}
B_{i}(t+1)=B_{i}(t)-D_{i}(t)+C_{i}(t)
\label{e0}
\end{equation}

To be specific, the energy storage has a lot of constraints. Firstly, we do not allow charging and discharging happen simultaneously:
\begin{equation}
1_{C_{i}(t)>0}+1_{D_{i}(t)>0}\leq{1}
\end{equation}
\begin{center}
$1_{f(x)>0}=\left\{\begin{array}{cc}
1 & \mbox{if}\ f(x)>0\\
0 & \mbox{otherwise}  \end{array}\right.$
\end{center}

Energy storage has finite capacity:
\begin{equation}
0\leq B_{i}(t) \leq B_{i}^{\max}
\end{equation}
where $B_{i}^{\max}$ is energy storage capacity. There are maximum charging rate $C^{\max}_{i}$ and discharging rate $D^{\max}_{i}$. Energy charging and discharging have to follow energy storage safety constraints:
\begin{equation}
0\leq{C_{i}(t)}\leq{\min[B^{\max}_{i}-B_{i}(t),C_{i}^{\max}(t)]}
\end{equation}
\begin{equation}
0\leq{D_{i}(t)}\leq{\min[B_{i}(t),D_{i}^{\max}(t)]}
\end{equation}

\begin{supertabular}{cp{65mm}}
\multicolumn{2}{c}{TABLE I}\\
\multicolumn{2}{p{75mm}}{THE INPUT QUANTITIES AND INTERMEDIATE VARIABLES}\\\hline
$R_{i}(t)$  &Amount of energy harvest by $MG_{i}$ in time slot $t$.\\
$G_{i}(t)$  &Amount of energy purchased by $MG_{i}$ from macrogrid in time slot $t$.\\
$X^{J}_{i}(t)$  &Energy purchased from auction for $MG_{i}$ in time slot $t$.\\
$P_{i}(t)$  &Price pay for per unit power from macrogrid, slot $t$\\
$I_{i}(t)$  &The delay-intolerant load demands arrived at the $MG_{i}$, slot $t$\\
$T_{i}(t)$  &The delay-tolerant load demands arrived at the $MG_{i}$, slot $t$\\
$B_{i}(t)$  &The amount of power stored in the battery of $MG_{i}$, slot $t$\\
$D_{i}(t)$  &Amount of energy discharged from $MG_{i}$ energy storage in time slot $t$.\\
$C_{i}(t)$  &The power charging to the battery at the $MG_{i}$, slot $t$\\
$J_{i}(t)$  &The power serve delay-tolerant load at the $MG_{i}$, slot $t$\\
$Q_{i}(t)$  &Length of queue buffering of delay-tolerant jobs at $MG_{i}$, slot $t$\\
$Z_{i}(t)$  &Length of virtual queue at $MG_{i}$, slot $t$\\
$\varepsilon_{i}$  &Set by users denote as user delay aware coefficient\\ \hline
\end{supertabular}\\

\subsubsection{Load Service}
The users of $\mbox{MG}_{i}$ have delay-intolerant (DI) $I_{i}(t)$ and delay-tolerant (DT) load demands $T_{i}(t)$. DI load demands need to be served when they come such as lighting. DT load demands should be served before a certain deadline like using washing machine and dish-wisher. In addition, we assume that both $I_{i}(t)$ and $T_{i}(t)$ are an independent and identically distributed random (i.i.d) non-negative stochastic process, and $0\leq T_{i}(t) \leq T_{i}^{\max}$. Now we define the $Q_{i}(t)$ as the length of queue buffering of delay-tolerant jobs at  $\mbox{MG}_{i}$  on time \textit{t} and $J_{i}(t)$ as the energy including both energy generation and purchase allocated to serve the DT load demands, then $Q_{i}(t)$ is according to following equation:
\begin{equation}
Q_{i}(t+1)=\max[Q_{i}(t)-J_{i}(t),0]+T_{i}(t)
\label{e1}
\end{equation}

The DT load demands can be delayed, but users still feel uncomfortable when these demands cannot be served. So we denote $Z_{i}(t)$ as delay aware queue as follow \cite{efficient} :
\begin{equation}
Z_{i}(t+1)=\max[Z_{i}(t)-J_{i}(t),0]+\varepsilon_{i}1_{Q_{i}(t)>0}
\end{equation}

No matter how, harvested energy must exceed energy demands, then we have:
\begin{equation}
\begin{aligned}
I_{i}(t)+J_{i}(t)+\sum^{N}_{l=1,l\neq{i}}\hat{y}_{il}(t)+C_{i}(t)\leq  R_{i}(t)+\\G_{i}(t)+D_{i}(t)+\sum^{N}_{l=1,l\neq{i}}\hat{x}_{il}(t)
\label{e8}
\end{aligned}
\end{equation}
where $x_{ij}(t)$ is the amount of energy bought by $\mbox{MG}_{i}$ from $\mbox{MG}_{l}$, and $y_{ij}(t)$  is the amount of energy sold from $\mbox{MG}_{i}$ to $\mbox{MG}_{l}$ .
Since we want storage queue can be used in \Ly optimization technique, we formulate a virtual energy storage queue as follow:
\begin{equation}
X_{i}(t)=B_{i}(t)-\Theta_{i}-D_{i}^{\max}
\label{e4}
\end{equation}
where $\Theta_{i}$ will be specified later.\\

\begin{supertabular}{cp{65mm}}
\multicolumn{2}{c}{TABLE II}\\
\multicolumn{2}{p{75mm}}{THE VARIABLES ABOUT DOUBLE AUCTION}\\\hline
$\beta_{i}(t)$   &The price of buying per unit energy by $MG_{i}$, slot $t$\\
$\alpha_{i}(t)$  &The price of selling per unit energy from $MG_{i}$, slot $t$\\
$\hat{\beta}(t)$  &Actual price of buying per unit energy, slot $t$\\
$\hat{\alpha}(t)$  &Actual price of selling per unit energy, slot $t$\\
$x_{ij}(t)$  &The amount of energy bought by the $MG_{i}$ to $MG_{l}$, slot $t$\\
$y_{ij}(t)$  &The amount of energy sold from $MG_{i}$ to $MG_{l}$, slot $t$\\
$\hat{x}_{ij}(t)$  &The actual amount of energy bought by the $MG_{i}$ to $MG_{l}$, slot $t$\\
$\hat{y}_{ij}(t)$  &The actual amount of energy sold from $MG_{i}$ to $MG_{l}$, slot $t$\\
$\beta_{i}^{min}$  &The minimum price that $MG_{i}$ can get a unit of energy. It should cover all necessary infrastructure costs\\
$\rho_{1},\rho_{2}$             &Set by Auctioneer measure the maximum energy can trade in the auction\\\hline
\end{supertabular}

\subsection{Problem Formulation}
\subsubsection{Microgrid Time Average Cost Minimization Problem}
The MGs want to cost less, meanwhile they can serve more demands. Firstly, we define energy expenditure $U_{i}(t)$ as follow:
\begin{equation}
\begin{aligned}
&U_{i}(t)=P(t)G_{i}(t)+\sum^{N}_{l=1,l\neq{i}}\hat{\beta}(t)\hat{x}_{il}(t)\\
&-\sum^{N}_{l=1,l\neq{i}}\hat{\alpha}(t)\hat{y}_{il}(t)
\end{aligned}
\end{equation}
where $\hat{\beta}(t)$ and $\hat{\alpha}(t)$ is actual price of selling and buying one unit of energy in double auction.

Then the time average cost minimization problem is shown
below:
\begin{equation}
\begin{aligned}
\bf{P1}: \, \min \quad \lim_{T\rightarrow\infty}\frac{1}{T}\sum_{t=0}^{T-1}\mathbb{E}\{U_{i}(t)\}
\end{aligned}
\end{equation}
subject to:
\begin{center}
\quad (\ref{e0})-(\ref{e4})
\end{center}
where the expectation is with regard to the random process in the system, control action and energy price. During each time slot, the decision variables are $C_{i}(t),\,D_{i}(t),\,J_{i}(t),\,G_{i}(t)$ and $\alpha_{i}(t),\,\beta(t),\,y_{il}(t),\,x_{il}(t)\,(\forall j \neq i)$.

\subsubsection{Double Auction Framework}
We want to use double auction to motivate MGs to take part in the electricity market.
We show the social welfare maximization problem \textbf{P2}:
\begin{equation}
\bf{P2:} \, \max \, \sum^{i'}_{i=1}\sum^{l'}_{l=1}(\rho_{1} \beta_{i'}(t)log(x_{il}(t))-\rho_{2}\alpha_{l'}(t)\frac{y_{li}^2(t)}{2})
\end{equation}
subject to:
\begin{equation}
\begin{aligned}
x_{il}(t)&=y_{li}(t) \\ \forall{i} \in \{1,2,\cdots,i' \}&,\,\forall{l} \in \{1,2,\cdots,l' \}
\label{e2}
\end{aligned}
\end{equation}
\begin{equation}
P_{i}(t) \geq \beta_{i'}(t) > \alpha_{l'}(t)
\label{e3}
\end{equation}
where $\beta_{i'}(t)$ and $\alpha_{l'}(t)$ are the sell-bid and buy-bid which may get accepted at time $t$. Constraint (\ref{e2}) makes sure the energy balance in the electricity market and both MGs and macrogrid can benefit from auction due to (\ref{e3}). The decision variables are $\beta_{i'}(t)$ and $\alpha_{l'}(t)$\\
\begin{figure}
\includegraphics[totalheight=56mm,width=80mm]{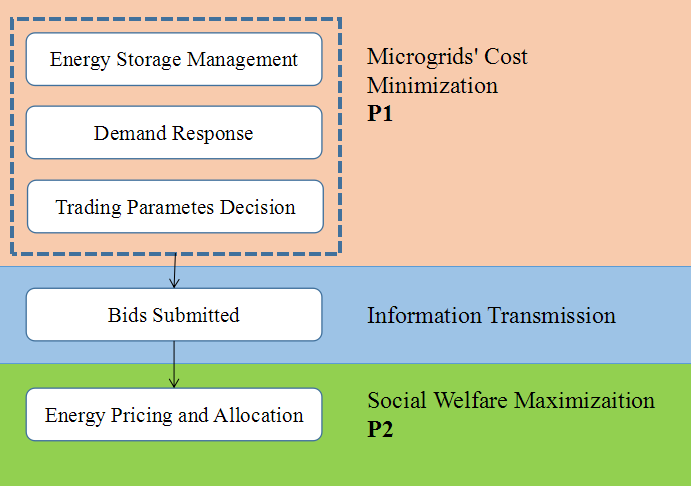}
\caption{The decision making process}
\end{figure}

\subsubsection{Discussion of the System Model}
We now illustrate some notice on the system model.
\begin{itemize}
\item We assume the energy bought by MGs from auction only can serve their load demands. In other words, the energy got from auction cannpot be charged into the energy storage and MGs cannot use energy bought from auction for further trading. In practice, we need consider about the efficiency of charging and discharging energy. If we put the charging and discharging efficiency into the system model, we will find that the possibility of purchasing energy for further trading is little. This more complicated model with physical model will be considered in the further work.
\item Transferring energy between the MGs will cause energy loss. However, the energy loss can be considered in our model which can be interpreted as a higher price which can cover the energy loss. Moreover MGs may use DC power line connection in the future, it will cost less energy than the traditional methods. So we ignore the energy loss caused by energy exchanging \cite{and}.
\item All MGs in our system model are selfish and rational. It means they only want to maximize their own profit and minimize their cost. They also want to fulfill customer demands as much as possible.
\end{itemize}

\section{Algorithm}
We next present a \Ly optimization based algorithm to solve \textbf{P1} and get optimization problem \textbf{P3} in Part A. Then we show this algorithm theoretical analysis and MG energy valuation and bids in Part B. In Part C, we design a double auction-mechanism for MG trading, which is economic efficient, individual rational, balanced budge and truthfulness and reformulate the \textbf{P2} into \textbf{P4}.
\subsection{Lyapunov optimization based Problem1}
Firstly, we need to define the \Ly function as
\begin{center}
$L_{i}(t)=\frac{1}{2}(Q_{i}^2(t)+X_{i}^2(t)+Z_{i}^2(t))$
\end{center}
The one-slot conditional Lyapunov drift can be defined as
\begin{equation}
\bigtriangleup{(L_{i}(t))}=\mathbb{E} \{ L_{i}(t+1)- L_{i}(t) | \overrightarrow{K}_{i}(t) \}
\end{equation}
And devote a vector $\overrightarrow{K}_{i}(t)=(Q_{i}(t),X_{i}(t),Z_{i}(t))$
At last we can formulate \textbf{P3}

\begin{equation}
\begin{aligned}
\bf{P3:}\min \quad &X_{i}(t)(C_{i}(t)-D_{i}(t))\\&-J_{i}(t)(Q_{i}(t)+Z_{i}(t))+V_{i}U_{i}(t)\\
\mbox{subject to:} &\lim_{T\rightarrow\infty}\frac{1}{T}\sum_{t=0}^{T-1}\mathbb{E}\{C_{i}(t)\}\\&=\lim_{T\rightarrow\infty}\frac{1}{T}\sum_{t=0}^{T-1}\mathbb{E}\{D_{i}(t)\}\\
&0\leq{C_{i}(t)}\leq{C_{i}^{\max}(t)}\\
&0\leq{D_{i}(t)}\leq{D_{i}^{\max}(t)}\\
&I_{i}(t)+J_{i}(t)+\sum^{N}_{l=1,l\neq{i}}\hat{y}_{il}(t)+C_{i}(t)\leq R_{i}(t)\\&+G_{i}(t)+D_{i}(t)+\sum^{N}_{l=1,l\neq{i}}\hat{x}_{il}(t)
\end{aligned}
\label{e5}
\end{equation}
Then we can have following Lemma.\\\\
\textbf{Lemma 1}: Given $\bigtriangleup{(L_{i}(t))}$ shown in (16)\\
We can have:
\begin{equation}
\begin{split}
&\bigtriangleup{(L_{i}(t))}+ V_{i}\mathbb{E}\{U_{i}(t)|\overrightarrow{K}_{i}(t)\} \\&\leq A_{i}+\mathbb{E}\{X_{i}(t)(C_{i}(t)-D_{i}(t))|\overrightarrow{K}_{i}(t)\}\\&+\mathbb{E}\{Q_{i}(t)(T_{i}(t)-J_{i}(t))|\overrightarrow{K}_{i}(t)\} \\&+\mathbb{E}\{Z_{i}(t)(\varepsilon_{i}(t)-J_{i}(t))|\overrightarrow{K}_{i}(t)\}\\&+V_{i}(\mathbb{E}\{U_{i}(t)|\overrightarrow{K}_{i}(t)\} )
\label{e6}
\end{split}
\end{equation}
where $A_{i}$ is the constant i.e.,\\
$A_{i}=\frac{(\varepsilon_{i}^{max})^{2}+(J_{i}^{max})^{2}}{2}+\frac{max[(C_{i}^{max})^{2},(D_{i}^{max})^{2}]}{2}+\frac{(J_{i}^{max})^{2}+(T_{i}^{max})^{2}}{2} $\\
\textbf{Proof:} At first, we can get following equation according (16)
\begin{enumerate}
\item $\frac{Q_{i}^{2}(t+1)-Q_{i}^{2}(t)}{2} \leq \frac{J_{i}^{2}(t)+T_{i}^{2}(t)}{2} +Q_{i}(t)(T_{i}(t)-J_{i}(t)) \\\leq \frac{(J_{i}^{max})^{2}+(T_{i}^{max})^{2}}{2}+Q_{i}(t)(T_{i}(t)-J_{i}(t))$\\

\item $\frac{X_{i}^{2}(t+1)-X_{i}^{2}(t)}{2} \leq \frac{(C_{i}(t)-D_{i}(t))^{2}}{2}+X_{i}(t)(C_{i}(t)-D_{i}(t)) \leq \frac{max[(C_{i}^{max})^{2},(D_{i}^{max})^{2}]}{2}+X_{i}(t)(C_{i}(t)-D_{i}(t))$\\

\item $\frac{Z_{i}^{2}(t+1)-Z_{i}^{2}(t)}{2} \leq \frac{\varepsilon_{i}^{2}(t)+J_{i}^{2}(t)}{2}+Z_{i}(t)(\varepsilon_{i}(t)-J_{i}(t)) \\\leq \frac{(\varepsilon_{i}^{max})^{2}+(J_{i}^{max})^{2}}{2}+Z_{i}(t)(\varepsilon_{i}(t)-J_{i}(t))$\\
\end{enumerate}

Next we can get:

$\bigtriangleup{(L_{i}(t))}+ V_{i}\mathbb{E}\{U_{i}(t)|\overrightarrow{K}_{i}(t)\} \\\leq \frac{(\varepsilon_{i}^{max})^{2}+(J_{i}^{max})^{2}}{2}+\frac{max[(C_{i}^{max})^{2},(D_{i}^{max})^{2}]}{2}+\frac{(J_{i}^{max})^{2}+(T_{i}^{max})^{2}}{2} +\mathbb{E}\{X_{i}(t)(C_{i}(t)-D_{i}(t))|\overrightarrow{K}_{i}(t)\}\\+\mathbb{E}\{Q_{i}(t)(T_{i}(t)-J_{i}(t))|\overrightarrow{K}_{i}(t)\} \\+\mathbb{E}\{Z_{i}(t)(\varepsilon_{i}(t)-J_{i}(t))|\overrightarrow{K}_{i}(t)\}\\+V_{i}(\mathbb{E}\{U_{i}(t)|\overrightarrow{K}_{i}(t)\} )$ \\Then the (\ref{e6}) directly follow.\\
Because $A_{i}+\mathbb{E}\{Z_{i}(t)\varepsilon_{i}(t)+Q_{i}(t)T_{i}(t)\}$ cannot change when the slot t begin and we need to minimize the right-hand of (\ref{e6}) and .Then we reformulate the \textbf{P3},we can get \textbf{P4}
\begin {equation}
\begin{split}
\bf{P4:}min \quad &X_{i}(t)(C_{i}(t)-D_{i}(t))\\&-J_{i}(t)(Q_{i}(t)+Z_{i}(t))\\&+V_{i}U_{i}(t)
\end{split}
\end{equation}
We denote $C_{i}^{*}(t)$, $D_{i}^{*}(t)$, $J_{i}^{*}(t)$, $G_{i}^{*}(t)$ and $\alpha_{i}(t)$, $\beta_{i}(t)$, $R^{J}_{i}(t)$, $X^{J}_{i}(t)$ as solution corresponding to $(15)$. $R^{J}_{i}(t)$ is the total energy that MGs can sell in time slot $t$ and $X^{J}_{i}(t)$ is the energy that MGs want from energy trading in time slot $t$.

\textbf{Lemma 2}: If MGs are rational and truthful, then:
\begin{equation}
\alpha_{i}(t)=\frac{Q_{i}(t)+Z_{i}(t)}{V_{i}}
\label{e11}
\end{equation}
\begin{equation}
\beta_{i}(t)=\left\{\begin{array}{c c}
\frac{Q_{i}(t)+Z_{i}(t)}{V_{i}} & \frac{Q_{i}(t)+Z_{i}(t)}{V_{i}}\geq \beta_{i}^{\min} \\
\beta_{i}^{\min} & \mbox{otherwise} \end{array}\right.
\label{e12}
\end{equation}

Respectively, the true values of the amount of energy to buy and to sell at $\mbox{MG}_{i}$ are:
\begin{equation}
R^{J}_{i}(t)=\left\{\begin{array}{ll}
R_{i}(t)-I_{i}(t) & \mbox{if}\ \alpha_{i}(t)>\frac{Q_{i}(t)+Z_{i}(t)}{V_{i}}\\
0 & \mbox{otherwise} \end{array}\right.
\label{n1}
\end{equation}
\begin{equation}
X^{J}_{i}(t)=\left\{\begin{array}{ll}
J^{\max}_{i}-R_{i}(t) & \mbox{if}\ \beta_{i}(t)<\frac{Q_{i}(t)+Z_{i}(t)}{V_{i}}\\
0 & \mbox{otherwise}p \end{array}\right.
\label{n2}
\end{equation}

\textbf{Proof.} We prove the Lemma2 depending on different cases\\
\newcounter{Meth}
\begin{list}{\textbf{Case\enskip\arabic{Meth}.}}{\usecounter{Meth}}
\item The buy-bid win, but sell-bid doesn't. Then \textbf{P4} transform into\\
$\mathbb{E}\{X_{i}(t)(C_{i}(t)-D_{i}(t)T_{i})\}\-\mathbb{E}\{J_{i}(t)(Q_{i}(t)+Z_{i}(t)T_{i})\}+V_{i}\mathbb{E}\{P(t)G_{i}(t)+\sum^{N}_{l=1,l\neq{i}}\hat{\beta}(t)\hat{x}_{il}(t)\}
$\\
When $\frac{Q_{i}(t)+Z_{i}(t)}{V_{i}}\geq \beta_{i}^{min}$\\
If $ \beta_{i}(t) > \frac{Q_{i}(t)+Z_{i}(t)}{V_{i}}$, the MGs tend to decrease $J_{i}(t)$ which means decreasing $\sum^{N}_{l=1,l\neq{i}}\hat{x}_{il}(t)$ and minimize \textbf{P4}.\\
If $ \beta_{i}(t) < \frac{Q_{i}(t)+Z_{i}(t)}{V_{i}}$, the MGs tend to increase $J_{i}(t)$ which means increasing $\sum^{N}_{l=1,l\neq{i}}\hat{x}_{il}(t)$ and minimize \textbf{P4}.\\
If $ \beta_{i}(t) = \frac{Q_{i}(t)+Z_{i}(t)}{V_{i}}$, the cost of per-unit energy and desire of fulfilling DT load demands are balance. So $ \beta_{i}(t) = \frac{Q_{i}(t)+Z_{i}(t)}{V_{i}}$ can reveal the true desire of $MG_{i}$.
When $\frac{Q_{i}(t)+Z_{i}(t)}{V_{i}} < \beta_{i}^{min}$\\
Then $\beta_{i}=\beta_{i}^{min}$
\item The sell-bid win, but buy-bid doesn't. Similar to the former Case,we can know:$\alpha_{i}(t)=\frac{Q_{i}(t)+Z_{i}(T)}{V_{i}}$
\item Both sell-bid and buy-bid are accepted.\\
We have constraint $\beta_{i}(t)>\alpha_{i}(t)$, but as a rational MG, $\beta_{i}(t)<\alpha_{i}(t)$. So this case is contradicted and cannot happen in a well-designed Double Auction.
\item Neither sell-bid nor buy-bid is accepted.\\
Then \textbf{P4} turn into $\mathbb{E}\{X_{i}(t)(C_{i}(t)-D_{i}(t)T_{i})\}-\mathbb{E}\{J_{i}(t)(Q_{i}(t)+Z_{i}(t)T_{i})\}+V_{i}\mathbb{E}\{P(t)G_{i}(t)\}$ The sell-bid and buy-bid don't have effect.
\end{list}

MGs need more economic incentive  to sell their renewable energy when they buffer lots of demands . In addition, they are willing to get energy with a higher bids.

After auction, algorithm can solve the linear programming problem (18) and get $C_{i}^{*}(t),\,D_{i}^{*}(t),\,J_{i}^{*}(t),\,G_{i}^{*}(t)$.

\subsection{Algorithm Analysis}
In this section, we summarize the properties of our algorithm as follows:\\
\textbf{Theorem 1.} All $V_{i}$ in (\ref{e5}) should meet the constraints that $0<V_{i}<V_{i}^{max}$ for all $t \in \{0,1,2,\cdots\}$
\begin{equation}
V_{i}^{max}=\frac{B_{i}^{max}-T_{i}^{max}-\varepsilon_{i}^{max}}{P^{max}-P_{i}^{min}}
\end{equation}

Then our Lyapunov optimization based Problem1 has the following properities:
\begin{enumerate}
\item Both $Q_{i}(t)$ and $Z_{i}(t)$ are upper bounded by $Q_{i}^{max}$ and $Z_{i}^{max}$ at all slot t,where
\begin{equation} \quad Q^{max}_{i}=V_{i}P^{max}+T^{max}_{i} \end{equation}
\begin{equation} \quad Z^{max}_{i}=V_{i}P^{max}+\varepsilon^{max}_{i} \end{equation}\\
Further here we denote $Q_{i}(t)+Z_{i}(t)$ upper bound as $\Theta_{i}$
\begin{equation} \Theta_{i}=V_{i}P^{max}+T^{max}_{i}+\varepsilon^{max}_{i} \end{equation}
\item we denote the worst-case delay as $\delta^{max}_{i}$,where
\begin{equation}
\begin{aligned}
&\delta^{max}_{i}=\frac{Q_{i}^{max}+Z_{i}^{max}}{\varepsilon_{i}}\\&=\frac{2V_{i}P^{max}+T^{max}_{i}+\varepsilon^{max}_{i}}{\varepsilon_{i}} \end{aligned}
\end{equation}
\item According the definition of virtual queue $X_{i}(t)$
\begin{equation} -\Theta_{i}-D_{i}^{max} \leq X_{i}(t) \leq B_{i}^{max}-\Theta_{i}-D_{i}^{max} \end{equation}
\item If $\forall{i}$, $R_{i}(t)$, $I_{i}(t)$ and $T_{i}(t)$ are i.i.d, then time-average-cost is shown below
\begin{equation}
\begin{split}
&\lim_{T \rightarrow \infty}\frac{1}{T}\sum_{t=0}^{T-1}\mathbb{E}\{P(t)G_{i}(t)+\sum^{N}_{l=1,l\neq{i}}\hat{\beta}(t)\hat{x}_{il}(t)\\&-\sum^{N}_{l=1,l\neq{i}}\hat{\alpha}(t)\hat{y}_{il}(t)\}\leq P^{*}_{1}+\frac{A_{i}}{V_{i}}
\end{split}
 \end{equation}
\end{enumerate}
\textbf{Proof.}We set $Q_{i}(0),Z_{i}(0)$ and $X_{i}(0)$ satisfying all these constraints, then we use induction to prove equation above.
\begin{enumerate}
\item If $0<Q_{i}(t) \leq V_{i}P^{max}$, then $Q_{i}(t) \leq VP^{max}+T_{i}^{max}$\\
If $Q_{i}(t) \geq V_{i}P^{max}$, because $Z_{i}(t)\geq{0} ,\forall t\in [0,1,2,\cdots]$ and our target is to minimize the \textbf{P4}. We can find $-(Q_{i}(t)+Z_{i}(t))+V_{i}P^{max}\leq{0}$. It means no matter how, increasing $G^{J}_{i}(t)$ can still minimize \textbf{P4}. So we choose $G^{J}_{i}(t)=G^{J,max}_{i}$, then \textbf{P4} can be minimized. According (\ref{e1}), $Q_{i}(t+1) \leq V_{i}P^{max}+T_{i}^{max}-J_{i}^{max} \leq V_{i}P^{max}+T_{i}^{max}$\\
To sum up, $Q_{i}^{max}=V_{i}P^{max}+T_{i}^{max}$, $Z_{i}^{max}$ and $\Theta_{i}$ can be proven similarly.
\item Then we will prove the worst-case delay $\delta_{i}^{max}$ by contradiction\\
$Z_{i}(t+1) \geq Z_{i}(t)-J_{i}(t)+\varepsilon_{i} 1_{Q_{i}(t)>0}$\\
So $Z_{i}(t_{0}+1+\delta_{i}^{max})- Z_{i}(t_{0}) \geq - \sum_{t=t_{0}}^{t_{0}+\delta_{i}^{max}}J_{i}(t)+\varepsilon_{i} \delta_{i}^{max}$\\
 Due to $Z_{i}(t_{0}+1+\delta_{i}^{max})\leq Z_{i}^{max}$ and $Z_{i}(t_{0})>0$\\
 Rearrange the terms\\
 $\sum_{t=t_{0}}^{t_{0}+\delta_{i}^{max}}J_{i} \geq \varepsilon_{i} \delta_{i}^{max}-Z_{i}^{max}$\\
 We assume the DT load demands came in at slot $t_{0}$ and cannot be served at slot $t_{0}+1+\delta_{i}^{max}$, So $\sum_{t=t_{0}}^{t_{0}+\delta_{i}^{max}}J_{i} \leq Q_{i}^{max}$\\
 Then $Q_{i}^{max} \geq \varepsilon_{i} \delta_{i}^{max}-Z_{i}^{max}$, rearrange the equation and use contradiction. We can get:\\
 $\delta^{max}_{i}=\frac{Q_{i}^{max}+Z_{i}^{max}}{\varepsilon_{i}}=\frac{2V_{i}P^{max}+T^{max}_{i}+\varepsilon^{max}_{i}}{\varepsilon_{i}} $
 \item To be specify that $\beta^{min}_{i}$ as the minimum price that $MG_{i}$ can get a unit of energy.\\
 When t=0, $X_{i}(0)=B_{i}(0)-\Theta_{i}-D_{i}^{max}\\<B_{i}^{max}-\Theta_{i}-D_{i}^{max}$\\
 If $B_{i}^{max}-\Theta_{i}-D_{i}^{max} \geq X_{i}(t)>0$, according to the \textbf{P4}, $D^{J}_{i}(t)>0,C_{i}(t)=0$. Then $B_{i}^{max}-\Theta_{i}-D_{i}^{max} \geq X_{i}(t)> X_{i}(t+1)$\\
 If $0 \geq X_{i}(t) > -V_{i}P^{min}$, then
 \begin{center}
 $-\Theta_{i}-D^{max}_{i}<-VP^{min}-D^{max}_{i}<X_{i}(t+1)<X_{i}(t)+C_{i}^{max}<B_{i}^{max}-\Theta_{i}-D_{i}^{max}$
\end{center}
Here we use the condition:\begin{center}$V_{i}<\frac{B_{i}^{max}-T_{i}^{max}-\varepsilon_{i}^{max}}{P^{max}-P^{min}_{i}}$\end{center}
If $-V_{i}P^{max} \geq X_{i}(t) > -\Theta_{i}-D_{i}^{max}$, according \textbf{P4}, $C_{i}>0,D_{i}=0$. Battery is charging. Then $X_{i}(t+1)>X_{i}(t) > -\Theta_{i}-D_{i}^{max}$\\
To sum up, $-\Theta-D_{i}^{max} \leq X_{i}(t) \leq B_{i}^{max}-\Theta-D_{i}^{max}$
\item $\sum^{T-1}_{t=0}V_{i}\mathbb{E}\{U_{i}(t)|\overrightarrow{K}_{i}(t)\} \leq V_{i}P_{1}^{*}T+\mathbb{E}\{L_{i}(T-1)\}-\mathbb{E}\{L_{i}(0)\}+B_{i}T$\\
    Dividing both sides by $V_{i}T$, letting $T\rightarrow\infty$, and using the condition that $\mathbb{E}\{L_{i}(T-1)\}$ and $\mathbb{E}\{L_{i}(0)\}$ are finite. Then:
    \begin{center} $\lim_{T \rightarrow \infty}\frac{1}{T}\sum_{t=0}^{T-1}\mathbb{E}\{P(t)G_{i}(t)+\sum^{N}_{l=1,l\neq{i}}\hat{\beta}(t)\hat{x}_{il}(t)-\sum^{N}_{l=1,l\neq{i}}\hat{\alpha}(t)\hat{y}_{il}(t)\}\leq P^{*}_{1}+\frac{A_{i}}{V_{i}}$\end{center}
\end{enumerate}

Theorem 1 implies that time average individual MG energy cost can be made close to offline optimum by increasing the value of $V_{i}$. However, this results in increasing customers dissatisfaction index $Q_{i}^{max}$, $Z_{i}^{max}$, $\delta^{\max}_{i}$ and cost of energy storage capacity $B_{i}^{max}$.

\subsection{Double Auction}
Double Auction Mechanism has two steps:
\begin{itemize}
\item MGs decide the sell-bid and buy-bid. In addition, MGs should submit the amount of energy they need or supply.
\item Auctioneer calculates according to \textbf{P2} and announces the accepted sell-bid and buy-bid, and the amount of energy that MG gets from or offer to the electricity can be determined.
\end{itemize}

In Section 2.2, we illustrate \textbf{P2}. We show the double auction mechanism in this section.\\
The auctioneer sorts all received buy-bids from MGs in descending order and sell-bids from MGs in ascending order in the sell prices:\\
$\overline{\beta}_{1}(t)\geq \overline{\beta}_{2}(t)\geq \cdots \geq\overline{\beta}_{n}(t)$ and $\overline{\alpha}_{1}(t)\leq \overline{\alpha}_{2}(t)\leq \cdots \leq\overline{\alpha}_{n}(t)$\\
Next we reformulate \textbf{P2} into \textbf{P5} based on new conditions:
\begin{equation}
\bf{P5:}\,\max  \sum^{\overline{i}}_{i=1}\sum^{\overline{l}}_{l=1}(\rho_{1}\overline{\beta}_{\overline{i}}(t)log(\overline{x}_{il}(t))-\rho_{2}\overline{\alpha}_{\overline{l}}(t)\frac{\overline{y}_{li}^2(t)}{2})
\end{equation}
subject to:
\begin{equation}
\begin{aligned}
\overline{x}_{il}(t)&=\overline{y}_{li}(t) \\ \forall{i} \in \{1,2,\cdots,\overline{i} \}&,\forall{l} \in \{1,2,\cdots,\overline{l} \}
\label{e15}
\end{aligned}
\end{equation}
where $\rho{1},\rho{2}$ are set by auctioneer to control energy trading. In order to maximize \textbf{P5}, we can estimate $\overline{x}_{il}(t)=\overline{y}_{li}(t)$ is close to $ \sqrt{\frac{\rho_{1}\overline{\beta}_{\overline{i}}(t)}{\rho_{2}\overline{\alpha}_{\overline{l}}(t)}}$ .

We assume $\overline{\beta}_{i^{*}}(t)$ and $\overline{\alpha}_{l^{*}}(t)$ as the solution to the \textbf{P5}, then  $\overline{\beta}_{i^{*}-1}(t)$ and $\overline{\alpha}_{l^{*}-1}(t)$ will be accepted by the auctioneer. $\overline{\beta}_{i^{*}}(t)$ and $\overline{\alpha}_{l^{*}}(t)$ can be solved by Lagrangian mechanic \cite{auction}.

The actual values of the buy/sell prices for $\mbox{MG}_{i}$ can be derived as:
\begin{equation}
\hat{\beta}_{i}(t)=\left\{\begin{array}{cc}
\overline{\beta}_{i^{*}}(t) & \mbox{if}\ \beta_{i}(t)\ \mbox{wins}\\
0 & \mbox{otherwise}  \end{array}\right.
\end{equation}
and
\begin{equation}
\hat{\alpha}_{i}(t)=\left\{\begin{array}{cc}
\overline{\alpha}_{l^{*}}(t) & \mbox{if}\ \alpha_{i}(t)\ \mbox{wins}\\
0 & \mbox{otherwise}  \end{array}\right.
\end{equation}

The optimal amount of energy purchased from $\mbox{MG}_{l}$ by $\mbox{MG}_{i}$ is:
\begin{equation}
\hat{x}_{il}(t)=\left\{\begin{array}{cc}
w_{x}(t)\cdot x_{il}(t) & \mbox{if}\ bid\ \beta_{i}(t)\ \mbox{win}\\
0 & \mbox{otherwise}  \end{array}\right.
\end{equation}

The optimal amount of energy sold from $\mbox{MG}_{i}$ to $\mbox{MG}_{l}$ is:
\begin{equation}
\hat{y}_{il}(t)=\left\{\begin{array}{cc}
w_{y}(t)\cdot y_{il}(t) & \mbox{if}\ bid\ \alpha_{i}(t)\ \mbox{win}\\
0 & \mbox{otherwise}  \end{array}\right.
\end{equation}
where $w_{x}(t)$ and $w_{y}(t)$ are the coefficients which can ensure (\ref{e15}).\\
\\
\textbf{Theorem 3.} Using the mechanism presented above, all MGs will submit the sell-bids and buy-bids truthfully, or they will get lower profit by deviating from the true value of the buy and sell bids in (\ref{e11}) and (\ref{e12})\\
\textbf{Proof.} This proof is similar to Theorem$2$ in \cite{virtual}, which is omitted here for brevity.

\section{Numerical Results}
In this section, we present numerical results based on the data from the real world to examine our algorithm in the previous sections.
\subsection{Experimental Setup}
We consider a network of five MGs, namely MG1, MG2, MG3, MG4, MG5 and MG6. Each MG includes renewable energy generation, energy storage, delay-intolerant loads and delay-tolerant loads. There are two types MGs. All MGs have DT and DI loads which are i.i.d. First type of MG (indexed by i=1,2,3) loads take value from [100,200] kWh uniformly at random. Second type of microgird (indexed by i=4,5,6) loads take value from [200,400] kWh uniformly at random. For the renewable energy generation, we use hourly average wind speed data provided by the Alternative Energy Institute (AEI) \cite{Institute}. Specifically, we choose the scaling factors such that the average wind-driven energy production during one slot is about 200 kWh for type-1 MGs and 600 kWh for type-2 MGs. Detailed data are shown in Fig.3(a). For the price purchasing energy from macrogrids, we use hourly energy price provided by the Power Smart Pricing administered for Ameren Illinois data \cite{Illinois}. The detail statistics are shown in Fig.3(b). The total length of data is 120 hours. The maximum storage capacity of both types MGs is 3MW.\\
\begin{figure}
\subfloat[Power Generated by microgrids]{\includegraphics[totalheight=30mm,width=90mm]{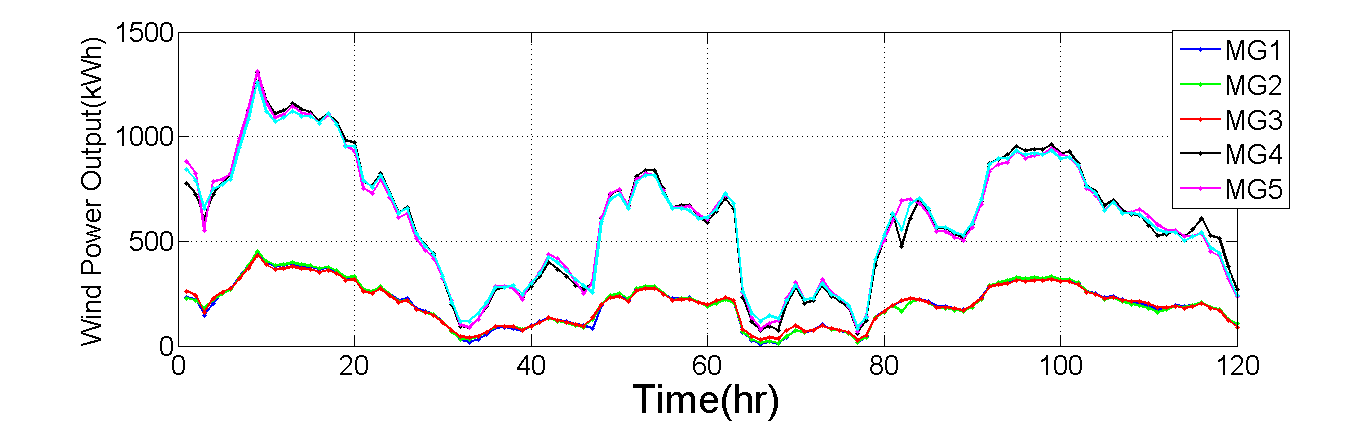}}\hspace{30pt}
\subfloat[Price for energy from macrogrid]{\includegraphics[totalheight=30mm,width=90mm]{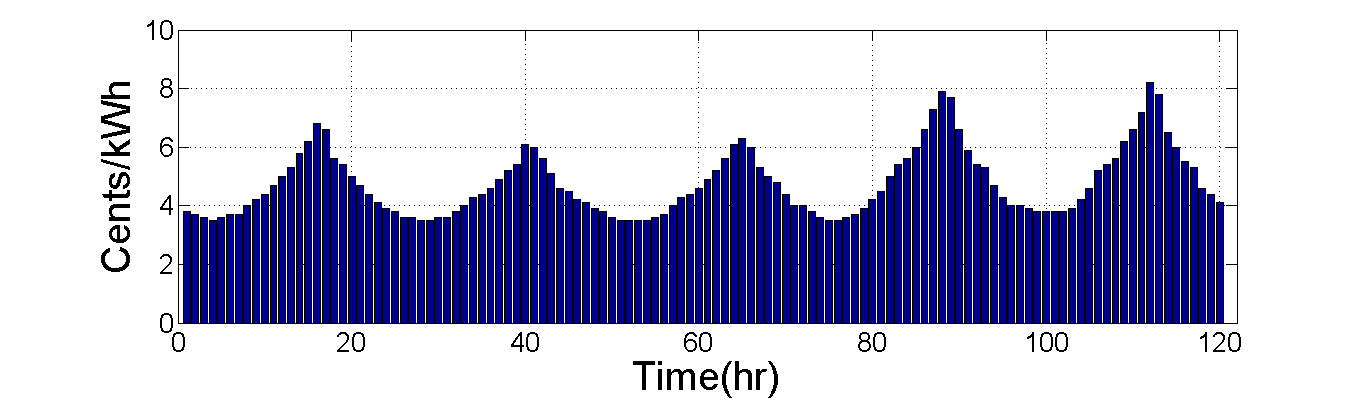}}
\caption{Data from Other Web-sites}
\end{figure}

In addition, the maximum charging and discharging rate of all MGs are 1.5MWh. Let $V_{i}=V_{i}^{\max}$ and $\varepsilon_{i}=T_{i}^{\min}$. At last, both type-1 and type-2 MG $\beta_{i}^{\min}=1$.

The auctioneer sets $\rho_{1}=1000$ and $\rho_{2}=0.0001$ which can make sure the most of bid can accepted in the auction.

\subsection{Results}
We show energy cost of all MGs on each time slot with our algorithm and without double auction mechanism in Fig.4. When MGs face renewable energy shortage like time slot $63-70$, double auction can reduce MG costs compared to individual MG operation. Although MG has to purchase extra energy from macrogrid to satisfy DT load demands after renewable energy shortage like time slot $40-50$, the average energy expenditure of all MGs are less than the situation without double auction and reduces about 11.3\%. In addition, our algorithm can ensure a certain delay of DT load demands.\\
\begin{figure}
\includegraphics[totalheight=30mm,width=90mm]{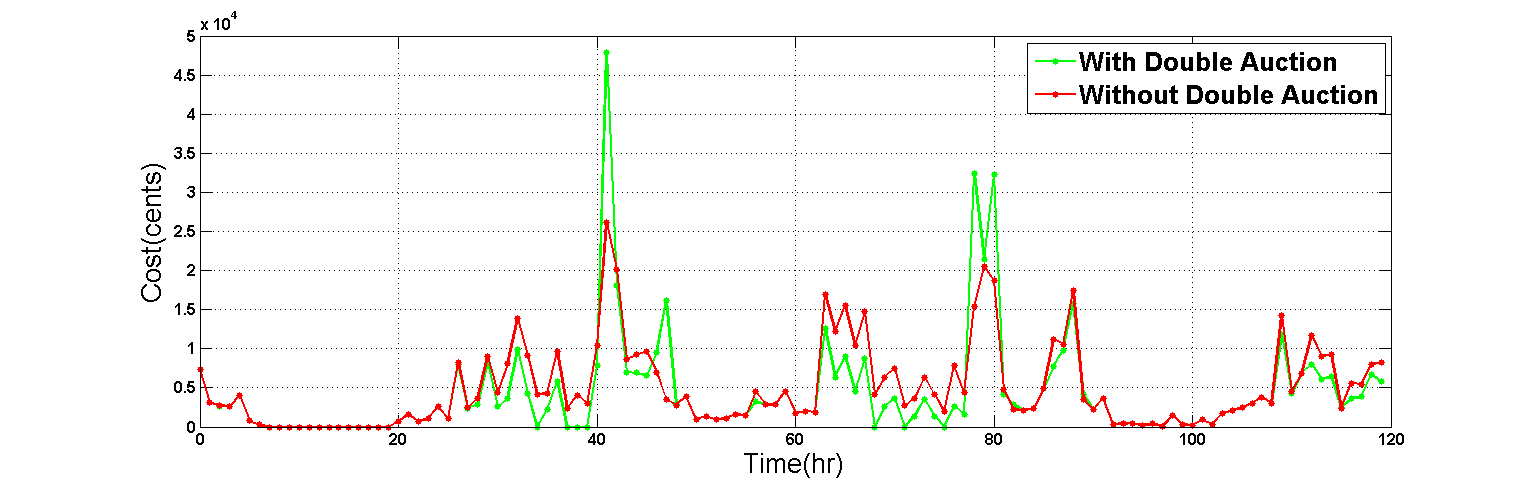}
\caption{Cost of all microgrids on each slot}
\end{figure}

Using double auction allow the MGs to exchange their extra renewable energy and purchase less energy from macrogrids. Fig.5 illustrates the energy purchased by all MGs from the macrogrids. In the most of slots, MGs using our algorithm need less energy from macrogrids and it outperforms the methods without double auction about 11.5\%.\\
\begin{figure}
\includegraphics[totalheight=30mm,width=90mm]{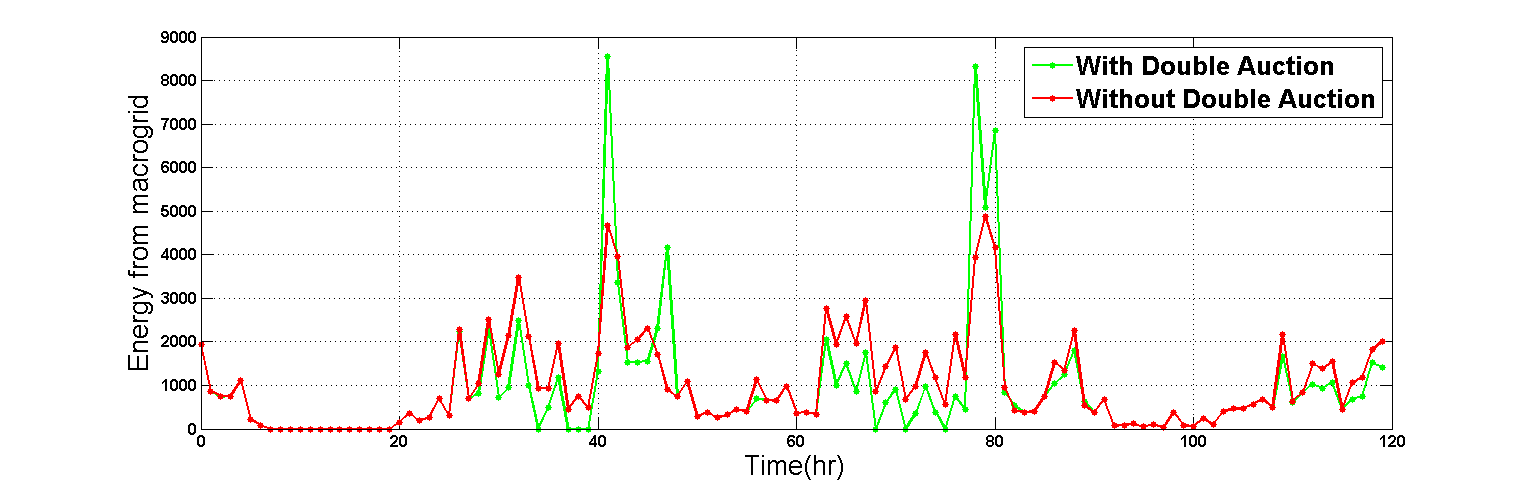}
\caption{Total energy purchased from macrogrid on each slot}
\end{figure}
\begin{figure}
\includegraphics[totalheight=30mm,width=90mm]{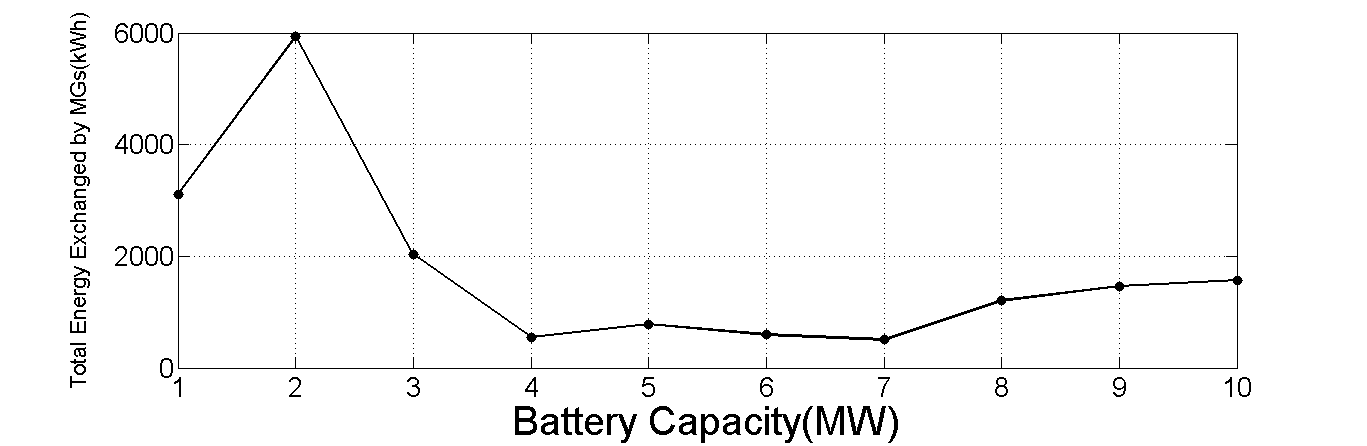}
\caption{Total energy exchanged between microgrids}
\end{figure}

At last, if MGs can not make some profits in the auction, they intend to store their extra energy into their energy storage. When $B_{i}^{\max}\rightarrow \infty$, MGs will seldom exchange extra energy. The double auction can motivate the MGs trading their energy into the markets. In Fig.6, even MGs have infinity capacity energy storage, they still will exchange energy.

\section{Conclusion}
This paper investigates both individual-cost minimization and social-welfare maximization strategies at individual-selfish MGs in energy trading market. We solve the time average cost minimization problem using \Ly optimization theory. Meanwhile, this algorithm can deal with stochastic problem brought by renewable energy generation and calculate trading bids. Then we propose a economic efficient, individual rational, balanced budge and truthful trading algorithm named double auction. At last, our results show that MGs energy expenditure can be reduced by our algorithm. Our solution can also be useful for MG designers to decide the capacity of energy storage and cooperation in order to meet certain energy expenditure criterion. In the future, we will take the energy transmission and energy charging efficiency into consideration. The analysis of how these constraints take effect on the energy trading and microgird energy management is left for future investigation.
\ifCLASSOPTIONcaptionsoff
  \newpage
\fi



%
\balance
\nocite{*}
\bibliographystyle{IEEEtran}

%

\end{document}